# Mesoscopic modeling and experimental validation of thermal and mechanical properties of polypropylene nanocomposites reinforced by graphene-based fillers

Atta Muhammad[a,b], Rajat Srivastava[a,c], Nikos Koutroumanis[d], Dionisis Semitekolos[e], Eliodoro Chiavazzo[a], Panagiotis-Nektarios Pappas[d], Costas Galiotis[d,f], Pietro Asinari[a,g], Costas A. Charitidis[e], Matteo Fasano[a*]

[a] *Department of Energy, Politecnico di Torino, Corso Duca degli Abruzzi 24, 10129, Torino, Italy.*

[b] *Department of Mechanical Engineering, Mehran University of Engineering and Technology, SZAB Campus, 66020 Khairpur Mir's, Sindh, Pakistan.*

[c] *Department of Engineering for Innovation, University of Salento, Piazza Tancredi 7, 73100, Lecce, Italy.*

[d] *Institute of Chemical Engineering Sciences, Foundation of Research and Technology-Hellas, Stadiou str. Rion, 26504, Patras, Greece*

[e] *School of Chemical Engineering, National Technical University of Athens, 9 Heroon Polytechniou, 15780 Athens, Greece*

[f] *Department of Chemical Engineering, University of Patras, 1 Caratheodory, 26504 Patras, Greece*

[g] *Istituto Nazionale di Ricerca Metrologica, Strada delle Cacce 91, 10135 Torino, Italy.*

[*] Corresponding author, email address: matteo.fasano@polito.it

**KEYWORDS:**

Nanocomposites; Multiscale modeling; Thermomechanical properties; Polymers; Graphene.

## ABSTRACT

The development of nanocomposites relies on structure-property relations, which necessitate multiscale modeling approaches. This study presents a modelling framework that exploits mesoscopic models to predict the thermal and mechanical properties of nanocomposites starting from their molecular structure. In detail, mesoscopic models of polypropylene (PP) and graphene based nanofillers (Graphene (Gr), Graphene Oxide (GO), and reduced Graphene Oxide (rGO)) are considered. The newly developed mesoscopic model for the PP/Gr nanocomposite provides mechanistic information on the thermal and mechanical properties at the filler-matrix interface, which can be then exploited to enhance the prediction accuracy of traditional continuum simulations by calibrating the thermal and mechanical properties of the filler-matrix interface. Once validated through a dedicated experimental campaign, this multiscale model demonstrates that with the modest addition of nanofillers (up to 2 wt.%), the Young's modulus and thermal conductivity show up to 35% and 25% enhancement, respectively, while the Poisson's ratio slightly decreases. Among the different combinations tested, PP/Gr nanocomposite shows the best mechanical properties, whereas PP/rGO demonstrates the best thermal conductivity. This validated mesoscopic model can contribute to the development of smart materials with enhanced mechanical and thermal properties based on polypropylene, especially for mechanical, energy storage, and sensing applications.

## 1. INTRODUCTION

In recent decades, polymers have been used in several industrial applications, ranging from the medical to the automotive industry, due to their light weight, corrosion resistance, low cost, and ease of manufacture.[1,2] Thermoplastics account for roughly 76% of the polymers used globally.[3]

The overall consumption of polypropylene (PP) is low compared to that of other thermoplastics such as polyethylene; however, its consumption has increased significantly in recent years owing to its interesting physical and chemical properties.[3] Polypropylene, an olefin, is partially non-polar and crystalline. It is fabricated from propylene monomer by chain-growth polymerization. The chemical formula for polypropylene is $(C_3H_6)_n$, and it is currently one of the low-priced polymers.[4] It can be processed through extrusion and injection molding.[4,5] Generally, polypropylene has a relatively low mechanical strength and poor thermal conductivity.[6,7]

Nanofillers can be included into the polypropylene matrix to form a composite with enhanced thermal and mechanical properties.[8,9] Polypropylene matrix can be reinforced with different nanofillers, such as glass fibers, aluminum oxide nanoparticles, and carbon nanotubes[10–12] to achieve desirable thermal and/or mechanical properties. For instance, Mirjalili *et al.*[11] performed morphological and mechanical characterization of polypropylene/nano $\alpha$-$Al_2O_3$ composites. They observed an increase in the elastic constant of PP by increasing the nano $\alpha$-$Al_2O_3$ content in the PP matrix from 1 to 4 wt.%. Further increase in the loading of nano $\alpha$-$Al_2O_3$ lead to a reduction in the elastic constant because of the agglomeration of $\alpha$-$Al_2O_3$ nanoparticles. Funck *et al.*[13] studied the multi-walled carbon nanotube (MWCNT)-reinforced polypropylene nanocomposites by in situ polymerization. Different MWCNT concentrations (0.1–8.0 wt.%) were introduced into the polypropylene matrix. They investigated their characteristics, such as morphology, crystallization and melting temperatures, and halftime of crystallization and melting temperature, and found that the halftime of crystallization decreases significantly as the filler content increases.

Given their superior properties, if introduced into the PP matrix, graphene-based nanofillers (e.g., graphene, graphene oxide, reduced graphene oxide) can significantly improve the material characteristics (such as elastic, thermal, electrical properties) even at small concentrations.[14–17]

Graphene (Gr) is a carbon allotrope comprising covalently linked carbon atoms bonded via $sp^2$ orbitals and structured in a two-dimensional hexagonal lattice.[18] Graphene oxide (GO) is a graphene derivative that has variable ratios of oxygen-rich functional groups on the basal plane and free edge, such as epoxide, carbonyl, carboxyl, and hydroxyl groups.[14] When compared to the pristine graphene, the presence of the functional group in GO weakens its in plane mechanical properties, such as elastic constant and intrinsic strength. GOs have been used in polymer nanocomposites because the thermal and electrical properties of polymers can be significantly enhanced by the incorporation of GO nanosheets.[19] Owing to its cost effectiveness and ease of production, reduced graphene oxide (rGO) is also commonly used as a filler with various materials to prepare nanocomposites. rGO is obtained from the reduction of graphene oxide using chemical, thermal, or photo-thermal reduction methods.[19] The fraction of the oxidized group is lowered when graphene oxide is reduced to obtain rGO, resulting in structural defects.[19] Nanocomposites with GO and rGO can be used for energy storage, stimuli responsive materials, anti-corrosion coatings, and separation applications.[19]

The technological development of nanocomposites with desired properties strongly depends on a deep understanding of the structure-process-property relationship with molecular precision.[20] Computer modeling is emerging as a powerful supplement to experimental and analytical approaches, which can provide improved understanding of the multiscale behavior of complex materials.[21,22] Multiscale modeling strategies provide seamless coupling among various length and timescales of materials properties and structures, from atomistic, to mesoscopic, then to continuum scale.[23] At the atomistic level, molecular dynamics (MD) simulations have been utilized to investigate the behavior of the nanocomposite constituent elements (such as polymer matrices and nanofillers) and their interaction at the interfaces.[24,25] The properties of the constituents of nanocomposites obtained at the atomistic level can be then employed in the mesoscopic model. The mesoscopic structure of a nanocomposite is represented by the representative volume element (RVE) of the material, and the properties computed at mesoscale can be finally homogenized to evaluate the effective thermal and mechanical properties at the macroscopic (continuum) level.[24]

Molecular dynamics simulations have allowed for the anticipation of material properties with well-parameterized and validated interatomic potential. Polymers have also been investigated using MD simulations with an accurate computation of their thermal and mechanical properties.[26,27,28] Wang *et al.*[29] performed MD simulation to observe the effects of molecular weight, chain number, and cooling rate on the glass transition temperature and coefficient of thermal expansion of polyethylene oxide. They found that the density increases with the chain length and thus molecular weight. Also, the glass transition temperature increased as the cooling rate increased, consistently with the experimental evidence.[30,31] Kim *et al.*[32] determined the hygroscopic and mechanical properties of semi-crystalline polypropylene using molecular dynamics and compared the results with experimental data. In particular, the elastic modulus and moisture saturation were studied with respect to the degree of crystallinity, concluding that the elastic modulus obtained from the MD simulation shows a similar trend to the experimental results and increases with degree of crystallinity. In case of moisture absorption, it was determined that a higher degree of crystallinity caused lower moisture uptake from both the MD simulation results and the experiments. Guryel *et al.*[33] used MD simulation to study the morphological and structural properties of three different polymeric nanocomposites reinforced with graphene. The polymers used in their study are polyethylene (PE), polystyrene (PS), and polyvinylidene fluoride (PVDF). They found the PE to have higher crystallinity than PVDF at a temperature of 500 K. Graphene influenced the crystallization of PVDF and PE because it acts as a nucleation site in both polymers.

Their results were in line with those obtained by previous quantum mechanical study.[34] Zhang *et al.*[35] also performed an MD simulation of glass fiber-reinforced polypropylene composites under various dynamic and thermal loadings. The interfacial strength decreases as temperature increases, resulting in a reduction in the mechanical properties of the matrix, whereas the mechanical properties increase with the strain rate.

Generally, atomistic models are computationally expensive and are limited to certain length and time scales. A possible solution to these problems is adopting mesoscopic coarse-grained (CG) models, which cluster groups of homogeneous atoms into one bead thus reducing the degrees of freedom of the system.[26,36] In recent years, the MARTINI coarse-grained model[37,38] has been an effective model for simulating polymers including polypropylene. In the MARTINI model, four heavy atoms and their accompanying hydrogens are represented by a single interaction center on average.[39] Panizon *et al.*[40] developed a CG model for polypropylene using structural and thermodynamic characteristics as earmark in the parametrization. As goal parameters, they considered densities and the radii of gyration for structural properties and segmentation of the various building blocks for thermodynamic properties, and the model was validated by matching the structural characteristics of the polymer. Ruiz *et al.*[41] developed a CG model of graphene based on the strain-energy conservation technique, where the model potentials are adjusted using the mechanical properties of graphene. The model can simulate mechanical responses in both the elastic and fracture domains. They found that the present model can be used for graphene-based nanocomposites. Similarly, Meng *et al.*[14] presented the CG model of graphene oxide using the strain energy conservation approach to optimize the potential parameters based on DFT calculations. They identified that the model could capture the mechanical and interfacial properties as well as the effect of oxidation in GO sheets, and hence is appropriate for inspecting the mechanical and interfacial properties of GO-based nanocomposites.

However, to the best of the Authors' knowledge, the properties of PP nanocomposites reinforced by graphene fillers have never been investigated by CG-MD. The present study proposes a new mesoscopic approach to determine the thermal and mechanical properties of polypropylene nanocomposites reinforced by graphene-based nanofillers. Initially, we determined the thermal and mechanical properties of pure polypropylene and graphene-based nanofillers (Gr, GO, and rGO). Then, the influence of the graphene-based nanofiller reinforcements on the thermal and mechanical properties of polypropylene was assessed. To compare the accuracy of the CG-MD model of PP/Gr nanocomposites with respect to traditional continuum approaches, finite element and mean field simulations were also carried out. Finally, an experimental thermo-mechanical characterization of polypropylene-graphene nanocomposites was conducted to validate the proposed multiscale modelling approach, where the peculiar properties of filler-matrix interface quantified by the mesoscopic model are employed to enhance the accuracy of continuum models.

## 2. METHODS

### *2.1 Mesoscopic models*

The considered coarse-grained model of polypropylene is taken from Panizon *et al.*,[40] who studied the interaction between PP and lipid membranes. They employed 3:1 mapping, as $CH_2$ groups are shared by neighboring CG beads as shown in Figure 1. The model contains bond, angle, and dihedral interactions. Harmonic functions describe the bonds and angles, whereas the sum of two proper dihedral functions describes the PP dihedrals, namely

$$V_b(r) = k_b(r - r_0)^2 \,, \qquad (1)$$

$$V_a(\theta) = k_\theta(\theta - \theta_0)^2 \,, \tag{2}$$

$$V_d(\phi) = k_\phi[1 + \cos(n\phi - \phi_s)] \,. \tag{3}$$

In Eq. 1, $r_0$ represents the equilibrium distance between the bonded beads, while $k_b$ the harmonic constant of their bond. In Eq. 2, $\theta_0$ is the equilibrium angle between a triplet of bonded beads and $k_\theta$ the angular harmonic constant. In Eq. 3, $k_\phi$, $n$ and $\phi_s$ are the parameters of the proper dihedral potential. The non-bonded interactions of PP are defined by the 12-6 Lennard-Jones potential:

$$V_{nb} = 4\varepsilon_{lj}\left[\left(\frac{\sigma_{lj}}{r}\right)^{12} - \left(\frac{\sigma_{lj}}{r}\right)^{6}\right], \tag{4}$$

being $\varepsilon_{lj}$ and $r_{lj}$ the energy well and equilibrium distance of the 12-6 Lennard-Jones potential between two non-bonded beads, respectively. The considered parameters of the CG force-field of PP are listed in the Supplementary Tables S1 and S2.

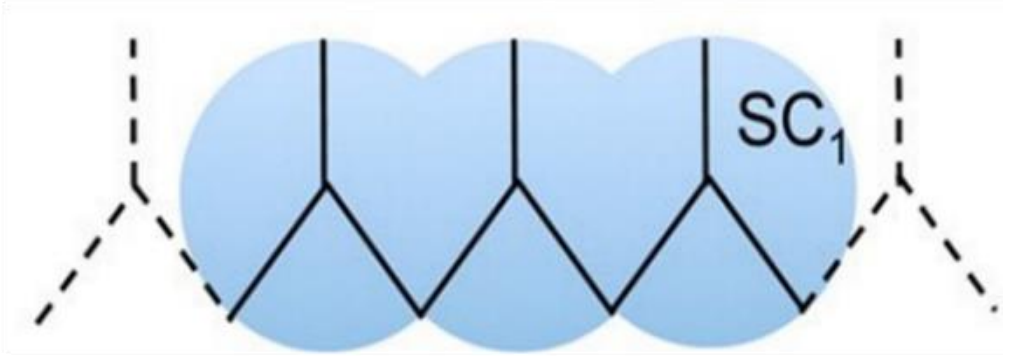

**Figure 1**. Coarse-grained representation of polypropylene (blue beads) from atomic details (black lines), where $SC_1$ are standard MARTINI beads. Reproduced from Ref. 40 with permission from ACS Publications.

The CG model of graphene, instead, is provided by Ruiz *et al.*[41] It follows the approach of Martini, where four atoms are clustered into a single bead as shown in Figure 2 (a) and preserves the hexagonal lattice of the beads.[42] The CG force field of the graphene model includes bonded and non-bonded interactions. Bonded interactions comprise bonds, angles, and dihedrals. On the one side, the bonding potential is now considered as

$$V_b(d) = D_0\left[1 - e^{-\alpha(d-d_0)}\right]^2, \quad (5)$$

where $D_0$ and $\alpha$ parameters are related to the depth and width of the potential well of the bond, respectively, $d_0$ represents the equilibrium distance of the bond. On the other side, Eqs. 2 and 3 are adopted to model angle and dihedral interactions, respectively. Non-bonded interactions are modelled by Eq. 4 as well. The considered parameters of the CG force-field of graphene are listed in the Supplementary Tables S3 and S4.

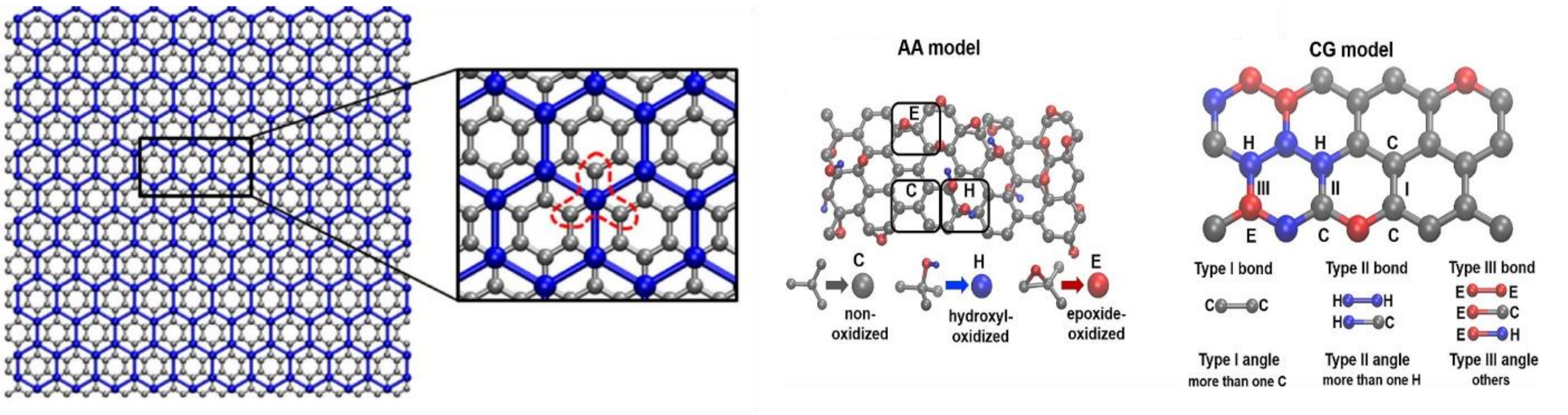

**(a)** **(b)**

**Figure 2**. (a) Coarse-grained model for graphene, reprinted from Ref. 41 with permission from Elsevier. (b) Coarse-grained model of graphene oxide, reprinted from Ref. 14 with permission from Elsevier.

Concerning GO, the CG model is inspired from Meng *et al.*,[14] who developed it for studying the mechanical behavior of graphene oxide. Like the Martini approach, 4:1 mapping scheme has been followed in GO, keeping the hexagonal structure of pure graphene also in this case. The CG model of GO includes the hydroxyl- and epoxide-oxidized functional groups as well as the non-oxidized group as shown in Figure 2 (b). The key characteristic that governs the mechanical behavior of GO is the degree of oxidation. Degree of oxidation is defined as the total percentage of both hydroxyl- (type H) and epoxide-oxidized (type E) beads. The force field of graphene oxide model includes bonds, angles, and non-bonded interactions. Accordingly, three bond and three angle types exist: non-oxidized, hydroxyl-oxidized, and epoxide-oxidized. The considered potential equations and related parameters of the CG force-field of GO are listed in the Supplementary Table S5. For the CG model of rGO, the same functional forms and parameters of GO have been utilized.

Coarse-grained MD simulations were run with a time step of 1 fs. The LAMMPS package[43] was used to perform the energy minimization and molecular dynamics calculation. The Nose-Hoover barostat[44] and thermostat[45] were used to control the pressure and temperature. The Verlet algorithm[46] was employed to integrate the equation of motions. VMD[47] was used to visualize the model and results. Initially, the models were annealed in an NVT ensemble at 500 K for 1 ns. After that, the simulation box was compressed in NPT ensemble at 5 atm and 300 K for 1 ns and then equilibrated in an NPT ensemble at 1 atm and 300 K for 10 ns. Finally, CG-MD simulations were run to determine the thermal and mechanical properties of the equilibrated models. The developed codes and numerical protocols are fully available at the Zenodo archive associated to this work.[48]

*2.2 Computation of material properties*

The glass transition temperature is a significant physical property of polymeric materials. It determines whether the polymeric material exhibits glassy- or rubbery-like behavior as well as the processing and working temperature range of the polymer. In the performed CG-MD simulations, the glass transition temperature is determined from the change in density or specific volume as a function of temperature at constant particle number, pressure, and temperature. This is because density and temperature have distinct linear relationship above and below glass transition temperature.[49]

Mechanical properties, such as Young's modulus and Poisson's ratio, of the modelled materials are also computed in the present study. In CG-MD simulation, uniaxial deformation is applied to the system, and the mechanical response of the system is recorded. Young's modulus of the equilibrated model is calculated from the uniaxial tensile test, while the Poisson's ratio is obtained using the theory of elasticity based on the Young's modulus. The deformation processes are carried out in three different directions, *x*, *y*, and *z*, at a temperature of 300 K. Polymers are typically isotropic materials, implying that material properties are constant in each direction. Isotropic materials have only two independent variables (elastic constants) in their stiffness and compliance matrices, whereas anisotropic materials may have up to 21 elastic constants. For the isotropic case, the two elastic constants are Young's modulus, *E*, and the Poisson's ratio, *v*, which are related as:

$$\begin{bmatrix} \varepsilon_{xx} \\ \varepsilon_{yy} \\ \varepsilon_{zz} \\ \varepsilon_{yz} \\ \varepsilon_{zx} \\ \varepsilon_{xy} \end{bmatrix} = \frac{1}{E} \begin{bmatrix} 1 & -\nu & -\nu & 0 & 0 & 0 \\ -\nu & 1 & -\nu & 0 & 0 & 0 \\ -\nu & -\nu & 1 & 0 & 0 & 0 \\ 0 & 0 & 0 & 1+\nu & 0 & 0 \\ 0 & 0 & 0 & 0 & 1+\nu & 0 \\ 0 & 0 & 0 & 0 & 0 & 1+\nu \end{bmatrix} \begin{bmatrix} \sigma_{xx} \\ \sigma_{yy} \\ \sigma_{zz} \\ \sigma_{yz} \\ \sigma_{zx} \\ \sigma_{xy} \end{bmatrix}, \quad (6)$$

where $\sigma$ is the stress vector and $\varepsilon$ the strain vector.

The Müller-Plathe method is used to investigate thermal conductivity (λ) of CG models. The method entails setting up two cold regions at the opposite ends of the simulation box. A certain amount of heat is applied in the central region (hot section), thus inducing a temperature gradient in the model. Velocities exchanged between the atoms of the hot and cold region generate heat flux. Periodic boundary conditions are applied in all three directions. Once the system reaches steady state, the amount of energy per unit time and cross-sectional area transferred from the hot region to the cold region via velocity exchanges between the molecules is:[50]

$$j_z = \frac{1}{2tA} \sum_{\text{transfers}} \frac{m}{2}\left(v_{\text{hot}}^2 - v_{\text{cold}}^2\right), \quad (7)$$

where $t$ is simulation time, $A$ is cross-sectional area normal to the heat flux direction, $m$ is mass, and $v_{hot}$ and $v_{cold}$ are the velocities of the defined atoms. The thermal conductivity can be then obtained using Fourier's law:[50]

$$\lambda = \frac{-j_z}{\nabla T}. \tag{8}$$

Finally, the specific heat capacity of the models at constant pressure ($c_p$) is also computed. In the CG-MD simulation, average enthalpies were recorded for every temperature step, and the $c_p$ of material determined from the slope of a linear fit of the resulting enthalpy-temperature plot.

### *2.3 Continuum models*

Two different continuum models, mean field (MF) and finite element method (FE), have been used in this study. MF homogenization is based on the first-order Mori-Tanaka, so it does not require RVE model and meshing. In FE, the RVE model is composed of PP matrix and graphene particulate (platelet like shape) inclusions with a completely bonded interface and an aspect ratio of 10 (like CG model). Periodic boundary conditions were applied to the RVE models.

### *2.4 Experimental methods*

We also investigated the experimental thermomechanical characteristics of pure PP, and PP samples filled with different graphene concentrations (0.5 and 1.0 wt.%) to validate the modelling results. To characterize the mechanical properties, a tensile test was performed. For the sample preparation, isotactic polypropylene was supplied by Sigma Aldrich, with average molecular weight *Mw* = 250,000, number of molecules *Mn* = 67,000, melt flow index (MFI) 12 g/10 min, and density 0.900 g/cm$^3$. The raw material for the nanofillers was graphite powder obtained from NGS Naturgraphit GmbH, Germany, with a particle average lateral size of 500 μm. The graphene was then produced using a commercially available shear laboratory mixer by Silverson. The mass of the produced exfoliated graphene was measured after drying at 80 °C for 24 h under vacuum conditions. The carbon content of produced graphene platelets was approximately 91%, with lateral dimensions in the range of 2–5 μm and a thickness of 5–7 nm. PP samples containing 0, 0.5, 1.0 wt.% of graphene were prepared using a hot press. For preparation of the polymeric film of PP, approximately 10 g of PP at 160 °C were heated using a hot press for about 5 min, and then pressed at a high pressure of 50 bar. The procedure was followed ten times to obtain a homogenous material. After heating and pressing, the film was taken out of the hot press, and then quenched in ice to obtain an amorphous structure and prevent crystallization. A similar procedure, with an additional step of melt (pre)mixing of nanoinclusions into the polymer matrix, has been adopted for the PP/Gr composite with two proportions: 0.5 and 1.0 wt.%.

Specimens of PP and PP/Gr were prepared following the international standard test method for tensile properties of polymers (ASTM D882-02). They were machined into $10 \times 1$ cm$^2$ rectangle specimens. The thicknesses of these samples were measured at various lengths, and the average was recorded; the area of each strip was calculated to determine the stress acting on the strips. Specimens were tested using a mechanical tensile system at 300 K and 1 atm at a strain rate of 25 mm/min. For statistical purposes, five samples of each composition were tested, and the average values of Young's modulus was obtained, as shown in Figure 3.

For investigating the thermal conductivity of pure PP and PP/Gr (0.5 and 1.0 wt.%) composites, test specimens with 30 mm length, 30 mm width, and 5 mm thickness were taken. Two samples of each composition were prepared. The measurements were performed five times for each sample using the transient plane source (TPS) method and the hot disk thermal constants analyzer instrument. Thermal properties of the polymer samples were examined according to the international standard ISO 22007 for the TPS method and the hot disk thermal constants analyzer instrument.[51] The sensors are positioned between the plane surfaces of two sample pieces of the material being studied. The hot disk sensor is made up of a double spiral electrically conductive pattern etched out of a thin sheet of nickel. The basic principle of the system is to constantly supply power to an initially isothermal sample via a hot disk sensor, and then use the same sensor as a resistance thermometer to follow the consequent temperature rise throughout a specified heating period. The dynamic characteristics of the temperature rise, reflected in sensor resistance increments, were carefully recorded, and analyzed, allowing for the determination of both the thermal conductivity and thermal diffusivity from a single transient recording.

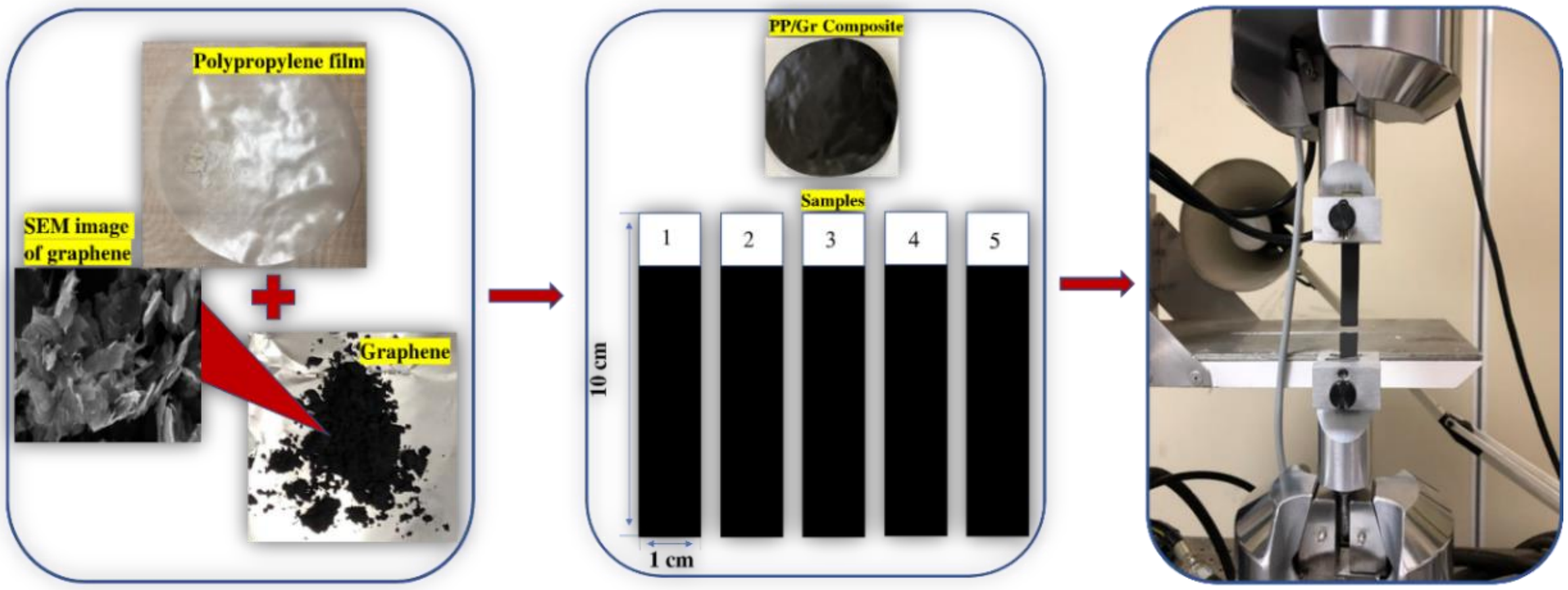

**Figure 3.** Mechanical characterization of PP/Gr composite.

## 3. RESULTS AND DISCUSSION

### *3.1 Polypropylene*

Boxes of 400 PP chains, with each chain containing 150 repetitive CG beads, as shown in Figure 4 (a), are equilibrated by performing NVT and NPT ensemble simulations for 10 ns. The density of PP computed at different temperatures (210–350 K) was found to be in the range of 1.13–1.17 g/cm$^3$, which is higher than the values reported in literature (0.90–0.91 g/cm$^3$).[52] This overestimation of density using MARTINI force field is also reported in previous studies.[40] In fact, the hydrophobic nature of PP suggests using the most hydrophobic MARTINI beads ($C_1$); however, since PP beads are connected by short bond length (0.29 nm), the smaller $SC_1$ beads are considered to represent each monomer. This choice leads to poor performance (overestimation) in terms of density.[40] Using densities at varied temperature levels obtained from NPT ensemble at a pressure of 1 atm, the glass transition was calculated by determining the intersection point between two fitting lines against density-temperature plot as shown in Figure 5 (a). The simulated glass transition temperature of PP ($T_g$ = 261–266 K) is in good agreement with data from previous studies ($T_g$ = 259 K).[53]

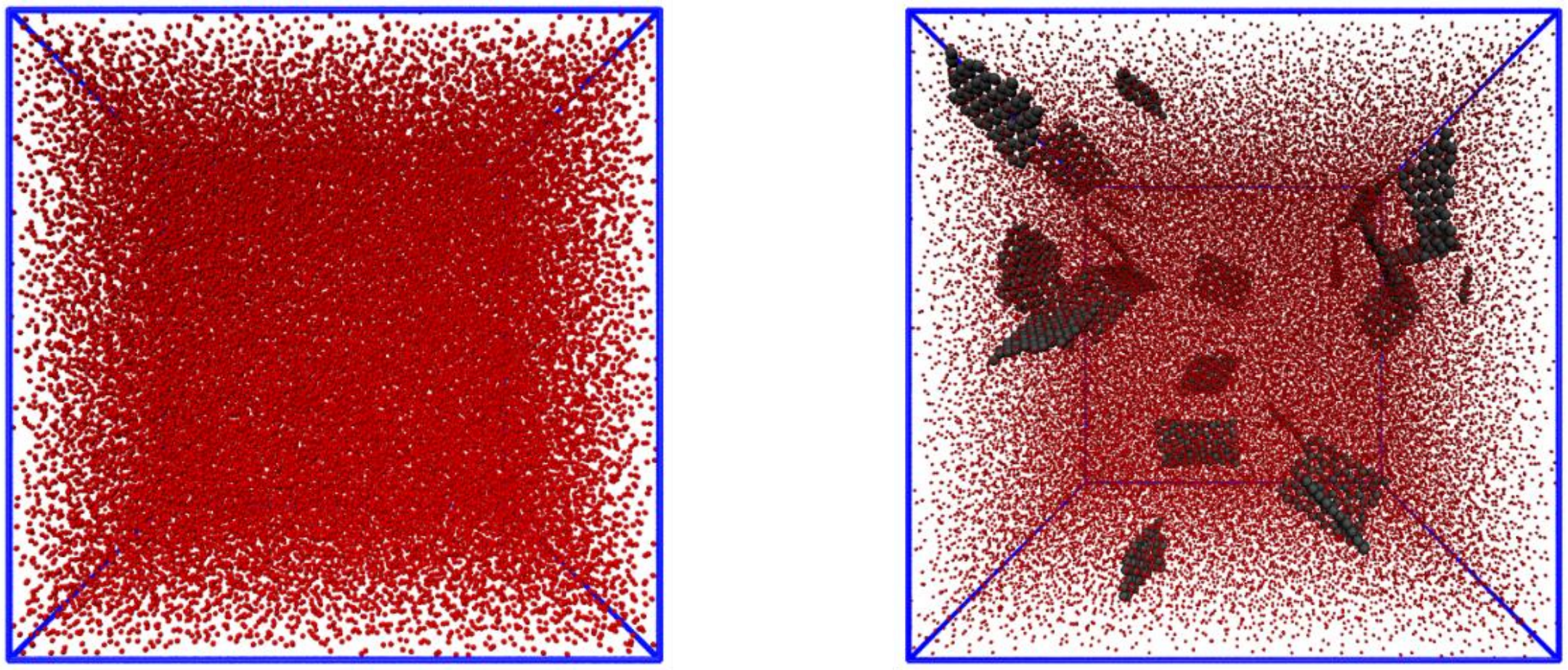

(a) (b)

**Figure 4.** Equilibrated CG model for (a) pure PP and (b) PP/Gr nanocomposite (2.0 wt.% Gr).

Mechanical properties such as the Young's modulus and Poisson's ratio are then extracted from stress-strain curves. The stress-strain curves obtained from the uniaxial tensile deformation in the *x*, *y* and *z* directions are shown in Figure 5 (b). The value of Young's modulus (0.99 GPa) is in line with our experimental result (0.94 GPa) and literature value (1.05 GPa).[54] Similarly, the computed Poisson's ratio (0.43) is in good agreement with the literature value (0.42).[55] Specific heat capacity of the PP was also computed from the enthalpy-temperature plot, as shown in Figure 5 (c). The best linear fit of this plot had a slope of 607 J/kgK, whereas the experimentally measured specific heat capacity of PP was 1,700 J/kgK at temperature of 300 K.[56] We also determined the thermal conductivity of neat polypropylene in all three directions. Results show that the average thermal conductivity of polypropylene at 300 K is 0.13 W/mK, which is lower than our experimentally measured value (0.23 W/mK) although in the range reported in literature (0.11–0.22 W/mK).[57,58]

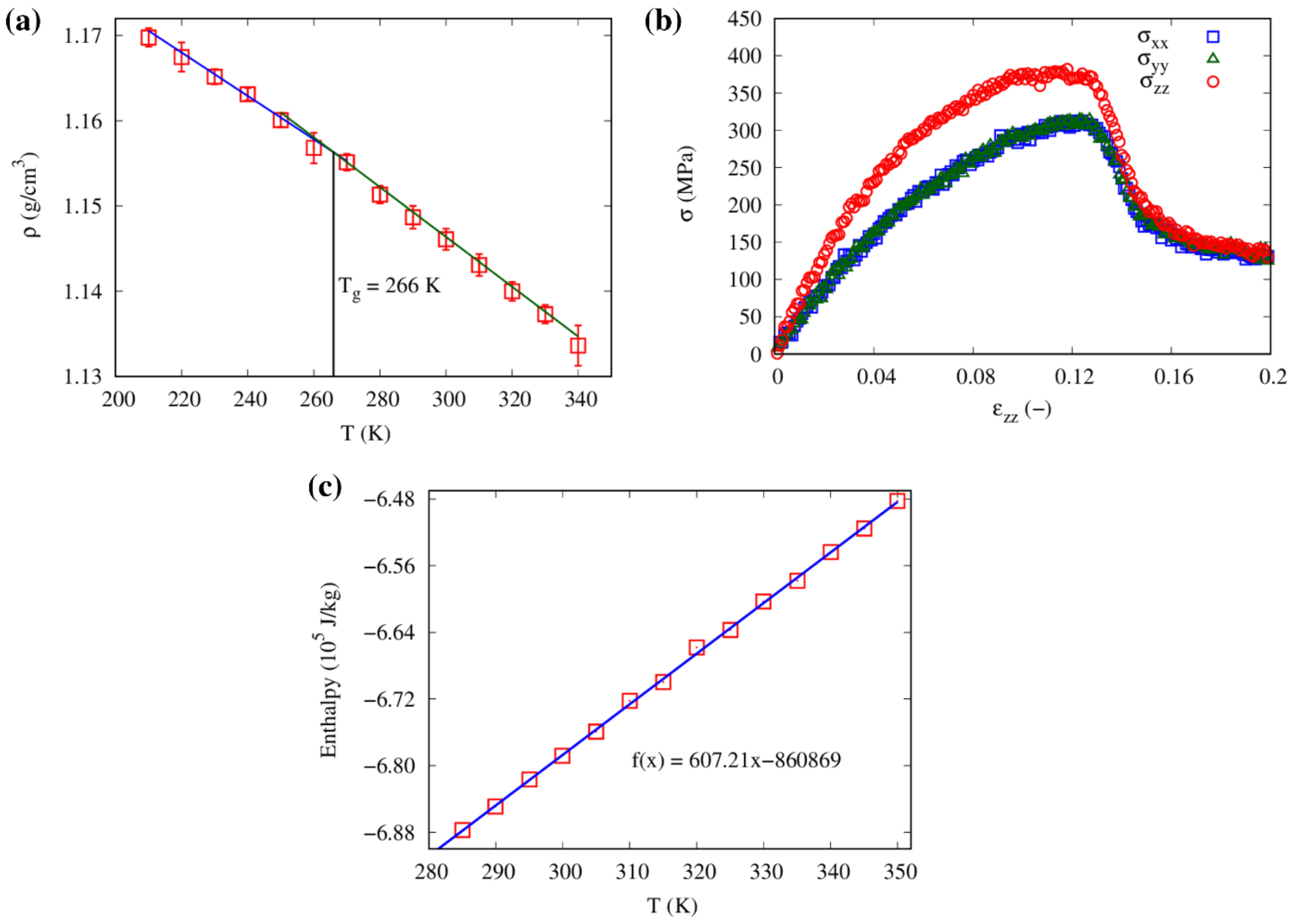

**Figure 5.** (a) Density as a function of temperature, (b) stress vs. strain curve and (c) enthalpy-temperature plot for the CG model of pristine polypropylene.

*3.2 Graphene-based fillers*

CG-MD simulations are then used to compute the thermal and mechanical properties of single layered graphene sheet in armchair and zigzag directions. The initial configuration of CG graphene was generated by VMD considering the bond length of the CG model of graphene. The considered size of graphene sheet was 20 × 40 nm$^2$. The $x$ and $y$ axis directions correspond to the zigzag and armchair edges, respectively. Initially, the system was equilibrated using NVT ensemble for 100 ps. After the equilibration, uniaxial tensile deformations along the armchair and zigzag directions were applied to the system, as shown in Figure 6 (a).

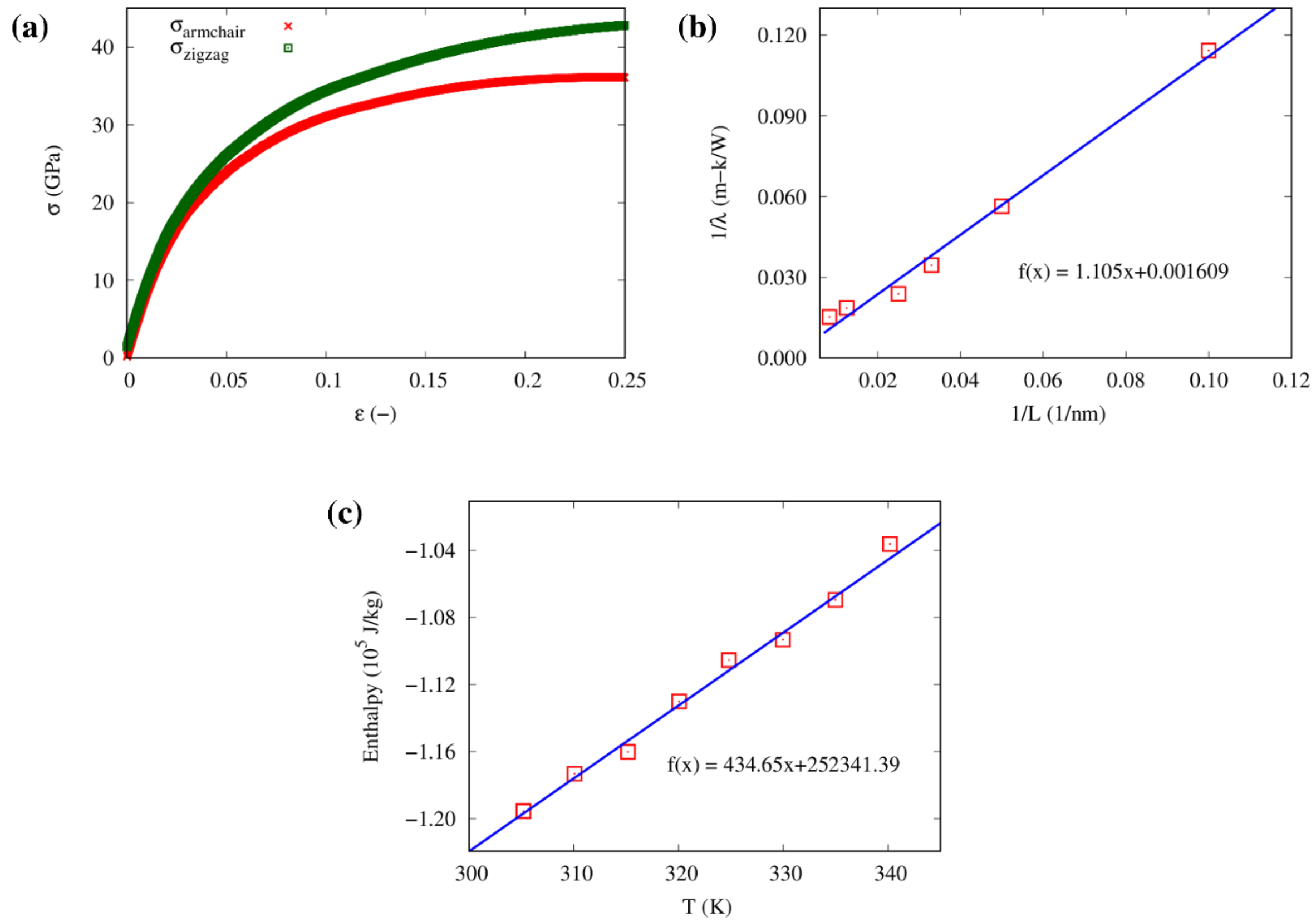

**Figure 6.** (a) Stress vs. strain response of single CG graphene sheet in armchair and zigzag direction, (b) inverse of thermal conductivity-inverse of length curve in armchair direction, and (c) enthalpy-temperature plot.

Young's modulus and Poisson's ratio of graphene in the armchair and zigzag directions were computed. Table 1 reports the comparison of the Young's modulus and Poisson's ratio in armchair and zigzag directions, showing good agreement with values reported in previous studies (900–1050 GPa and 0.14–0.19, respectively).[41,59] We also investigated the thermal conductivity of the CG graphene with different lengths using reverse non-equilibrium CG-MD simulations. We consider samples with lengths ranging from 10 nm to 100 nm. The dependence of the inverse thermal conductivity $1/\lambda$ to inverse length $1/L$ is illustrated in Figure 6 (b).

| **Direction** | **Young's modulus (GPa)** | **Poisson ratio (-)** | **Thermal conductivity-$\lambda_0$ (W/mK)** |
|---|---|---|---|
| **Armchair** | 845 | 0.14 | 621.5 |
| **Zigzag** | 916 | 0.15 | 658.7 |

**Table 1**. Mechanical and thermal properties of graphene as computed by CG-MD.

Generally, the relationship between length and thermal conductivity in graphene can be adequately described using the ballistic-to-diffusive crossover formula:[60]

$$\frac{1}{\lambda(L)} = \frac{1}{\lambda_0}\left(1 + \frac{\mathcal{L}}{L}\right). \quad (9)$$

Here, $L$ is the mean free path of phonon in graphene, and $\lambda_0$ is the thermal conductivity with infinite length. We make the linear fitting of CG-MD results as shown in Figure 6 (b). The simulated value of thermal conductivity at infinite length available in previous studies is 746 W/mK,[60] which is only slightly higher than the results from the tested mesoscopic model shown in Table 1. Specific heat capacity of the CG graphene filler was also computed from the enthalpy-temperature plot, as shown in Figure 6 (c). The best linear fit of this plot had a slope of 434 J/kgK, whereas the experimentally measured specific heat capacity of graphene was 700 J/kgK at temperature of 300

K.[61] Such discrepancy in the specific heat capacity is due to the reduction in degrees of freedom at the CG level, being an intrinsic limitation of mesoscopic models.[62,63]

We also evaluated the Young's modulus of the CG model of graphene oxide and reduced graphene oxide and compared them with literature evidence. We performed calculations for different degrees of oxidation of GO, and the results show that the Young's modulus decreases from 412 GPa to 287 GPa with increasing degree of oxidation as shown in Figure 7 (a), consistently with previous studies.[14] Note that equal ratio (1:1) of the hydroxyl-oxidized and epoxide-oxidized beads are considered in the model. Similarly, Figure 7 (b) shows the uniaxial tensile results of rGO with different percentage of defects. We constructed the CG model of rGO by randomly deleting the carbon beads of the CG GO model to generate defects. Note that the monolayer sheet of GO with 4% degree of oxidation was considered for construction of rGO sheet with different percentage of defects. We observed that the Young's modulus decreases from 316 GPa to 151 GPa as the percentage of defects in the rGO sheet increases.

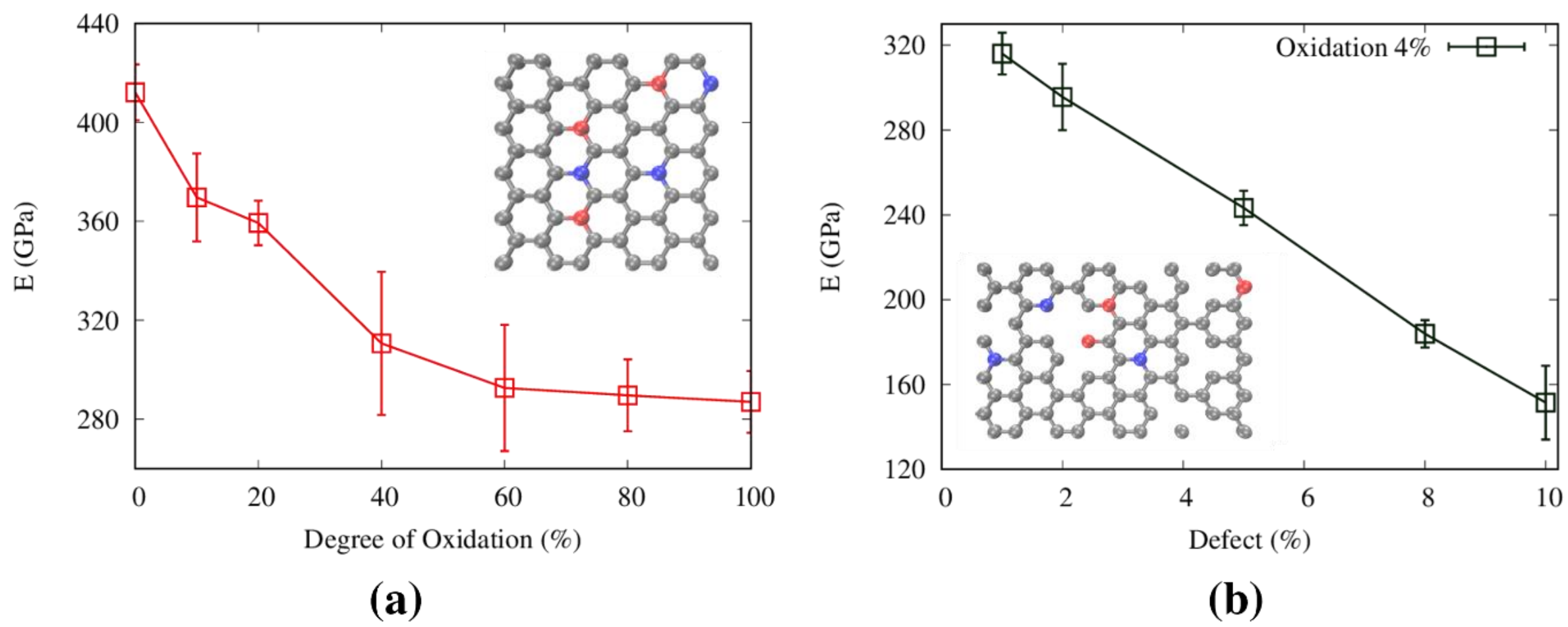

**Figure 7**. Average values and standard deviation of (a) Young's modulus vs. degree of oxidation for the CG model of GO and (b) Young's modulus vs. defect percentage for the CG model of rGO. Sampled were tested in both armchair and zigzag directions and results averaged.

*3.3 Polypropylene nanocomposites reinforced by graphene-based fillers*

We finally used CG-MD simulation to understand the influence of graphene-based inclusion on the thermal and mechanical properties of pure polypropylene. For PP/Gr, PP/GO, and PP/rGO composites, we randomly introduced PP chains and Gr/GO/rGO sheets into a simulation box, considering different nanofiller concentrations (0.5, 0.8, 1.0, 1.5, and 2.0 wt.%). The CG system was then energy minimized through NVT and NPT runs at a temperature of 300 K and pressure of 1 atm (time step of 1 fs; simulation time of 10 ns). The relaxed system was eventually used to determine the thermal and mechanical properties.

Figure 8 (a) shows the overall comparison of relative Young's moduli of PP/Gr, PP/GO, and PP/rGO nanocomposites. In the case of PP/Gr nanocomposites, the Young's modulus increases from 5.7% (0.5 wt.%) to 35.4% (2.0 wt.%), whereas it increases from 3.3% (0.5 wt.%) to 13.14% (2.0 wt.%) and from 0.8% (0.5 wt.%) to 10.71% (2.0 wt.%) for PP/GO and PP/rGO, respectively, compared to the pure value of PP. Overall, PP/Gr exhibits better mechanical properties than PP/GO and PP/rGO. In fact, the presence of functional groups in GO deteriorates the mechanical properties, as also observed in a previous study.[14]

Figure 8 (b) shows the comparison of relative thermal conductivities of PP/Gr, PP/GO, and PP/rGO nanocomposites. Thermal conductivity increases as the weight percentage of nanofillers increases. However, PP/GO and PP/rGO exhibit higher thermal conductivity than PP/Gr. The incorporation of GO into the polymer matrix could significantly improve the original thermal and electrical properties owing to the presence of functional groups leading to enhanced filler-matrix affinity, as also reported by Meng.[14] A detailed list of all CG-MD results is available in the Supplementary Table S6.

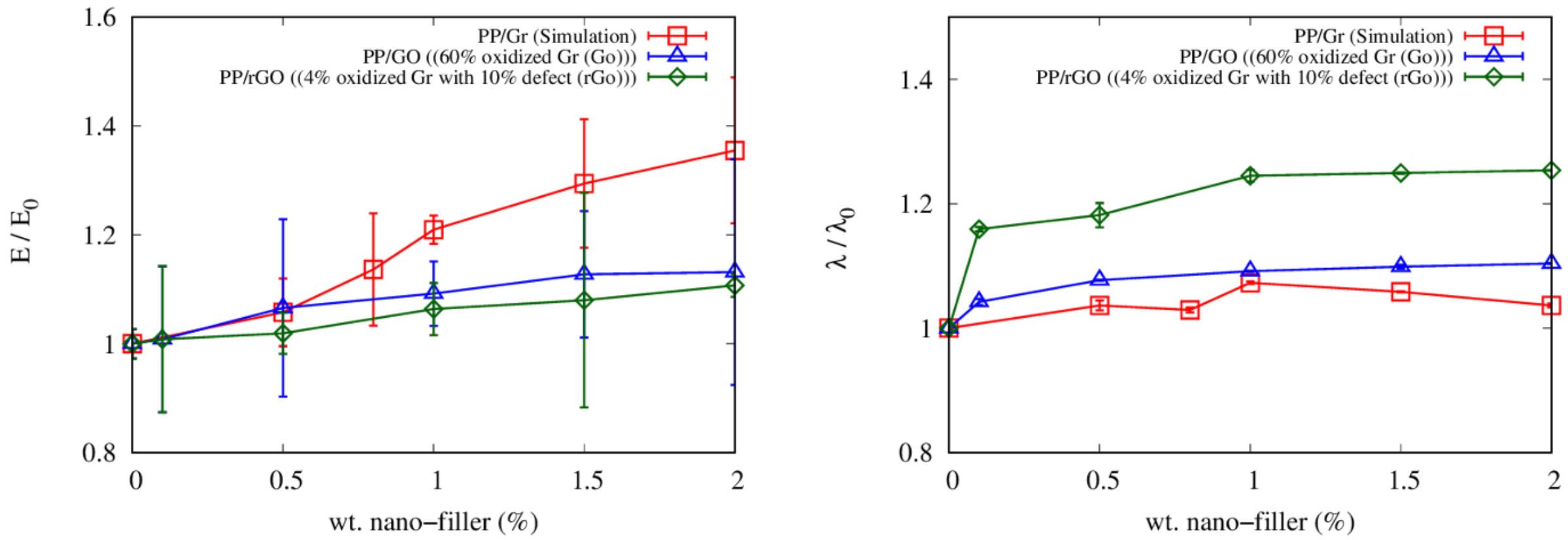

**(a)** **(b)**

**Figure 8.** Average values and standard deviation of relative (a) Young's modulus and (b) thermal conductivity with respect to the values of pristine PP ($E_0 = 0.989,$ and $\lambda_0 = 0.138$), respectively. Each sample was tested in $x$, $y$, and $z$ directions and results averaged.

### *3.4 Comparison and calibration of continuum approaches with mesoscopic results*

The finite element method is widely used to determine the thermal and mechanical properties of nanocomposites at the macroscopic (continuum) level. FE can be used to calculate the thermal conductivity of nanocomposites by solving Fourier's law for conduction.[24] It can also be employed to numerically evaluate the macroscopic mechanical properties of nanocomposites, such as the Young's modulus and Poisson's ratio.[24] For instance, Moghaddam *et al.* [25] studied composites with randomly distributed fillers (glass particles) in a polymer matrix (epoxy) by stochastic finite element analysis, exploring the Young's modulus, Poisson's ratio, coefficient of thermal expansion, and thermal conductivity. Peng *et al.*[64] proposed a numerical-analytical model for the nano-reinforced polymer composites and examined the microstructures and mechanical properties of the composites. Saber-Samandari *et al.*[65] used a finite element model to determine the elastic modulus of the nanocomposites with different inclusion shapes, such as platelet, spherical, and cylindrical ones. However, traditional continuum approaches cannot explicitly model the filler-matrix interactions, therefore being unable to represent the effect of different physical-chemical features of the interface. Mesoscopic simulations allow to overcome this issue since they model the filler-matrix interaction with molecular precision. Here, for the sake of completeness, we first compare the predictions from our CG model against traditional continuum predictions.

Hence, we used continuum models to evaluate the Young's modulus, Poisson's ratio, and thermal conductivity for the considered PP/Gr nanocomposites. The input parameters were taken coherently with CG models as follows: polypropylene – density 1.15 g/cm$^3$, Young's modulus 0.99 GPa, Poisson's ratio 0.43, and thermal conductivity 0.13 W/mK; graphene – density 1.40 g/cm$^3$, aspect ratio 10, Young's modulus 916 GPa, Poisson's ratio 0.15, and thermal conductivity 8.75 W/mK. The RVE model was then generated similarly to the CG configurations by choosing the mass fraction (0.5, 0.8, 1.0, 1.5) wt.% and thus the number of graphene inclusions dispersed randomly in the PP matrix, see Figure 9.

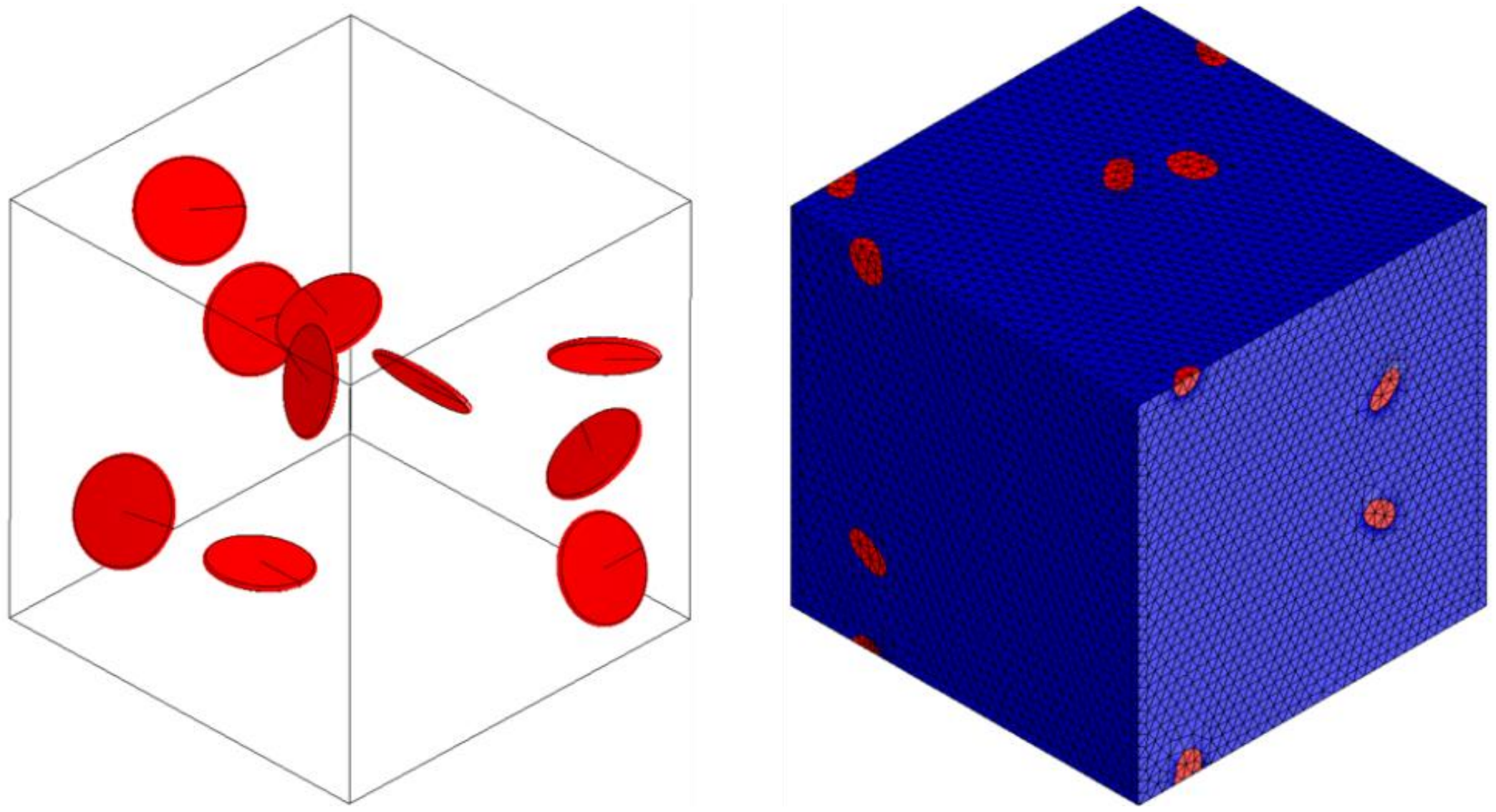

**Figure 9.** Example of computational domain and resulting mesh for the finite element model of polypropylene (blue) reinforced by graphene (red).

The results obtained from the FE and MF models are compared with the CG simulations in Figure 10. The Young's modulus increases from 5.7% (0.5 wt.%) to 35.4% (2.0 wt.%) compared with the neat PP in CG-MD simulation, while it increases from 2.2% (0.5 wt.%) to 8.0% (2.0 wt.%) with the FE and MF models (see Figure 10 (a)). Hence, both FE and MF models underestimate the Young's modulus enhancement provided by graphene inclusions, whereas the proposed CG model takes into consideration the filler-matrix interaction with molecular precision. The Poisson's ratio of PP/Gr composites, as predicted by the mesoscopic and continuum models, is shown in Figure 10 (b). The CG, FE and MF models all predict a progressive reduction of the Poisson's ratio with increasing graphene concentrations, with the CG simulation predicting higher reduction compared to both FE and MF models.

To consider the effect of matrix-filler interactions also in continuum approaches, Ji *et al.*[66] proposed an interphase micromechanical model based on the Takayanagi[67–69] homogenization technique, while accounting for interfacial contribution. Following the Ji *et al.* approach, we have introduced an interphase in the Mori-Tanaka MF model, considering a thickness of 0.5 nm from the fiber surface as observed in the mesoscopic simulations. Numerous studies[70–73] have revealed that the polymeric region in the proximity of the filler (i.e., interphase) exhibits significantly enhanced elastic modulus when transitioning from the nanoparticles to the polymer matrix. It is crucial to consider such interfacial stiffening, as neglecting it could lead to inaccurate predictions of the overall nanocomposites properties. Therefore, a comprehensive characterization of the interphase becomes essential to improve the prediction accuracy of continuum approaches able to upscale atomistic or mesoscopic models to the macroscopic sizes of experimental samples. To this purpose, the CG-MD results were taken as a reference to best fit the mechanical properties of the interphase included in the MF model, which resulted in a Young's modulus of 5 GPa and a Poisson's ratio of 0.44 clearly showing enhanced mechanical properties of the interphase. Results in Figures 10 (a) and (b) demonstrate a good agreement between the CG-MD and the calibrated interphase MF model, which is finally capable to accurately reproduce the effect of filler-matrix interface on the mechanical properties of the nanocomposite.

Figure 10 (c) compares the thermal conductivity of PP/Gr composites predicted by mesoscopic and continuum approaches. The thermal conductivity of the PP composite material increases with the weight percentage of graphene fillers in all cases, with CG-MD predictions showing lower enhancement since thermal boundary resistances at the filler-filler and filler-matrix interfaces are duly considered in this model – while being neglected by continuum ones.[74,75]

To include the effect of thermal boundary resistance also in continuum approaches, Balandin *et al.*[76] proposed a modified Maxwell-Garnett effective medium approximation incorporating an interphase with thermal properties dictated by the Kapitza resistance at the interface between epoxy and graphene. Following the methodology outlined by Balandin *et al.*, we have incorporated an interphase in the Mori-Tanaka MF model, adopting a thickness consistent with the mechanical model. Then, the Kapitza resistance of the interphase between polymer and graphene was taken as $3.7\times10^{-9}$ $m^2K/W$, coherently with previous observations in the literature.[75,76] This led to a thermal conductivity of the interface of about 0.135 W/mK, which is similar to the CG thermal conductivity of neat PP. The inclusion of thermal boundary resistance at the filler-matrix interface improves the agreement between the interphase MF model and the CG simulation result. A detailed list of all such model comparisons is available in the Supplementary Tables S7-S9.

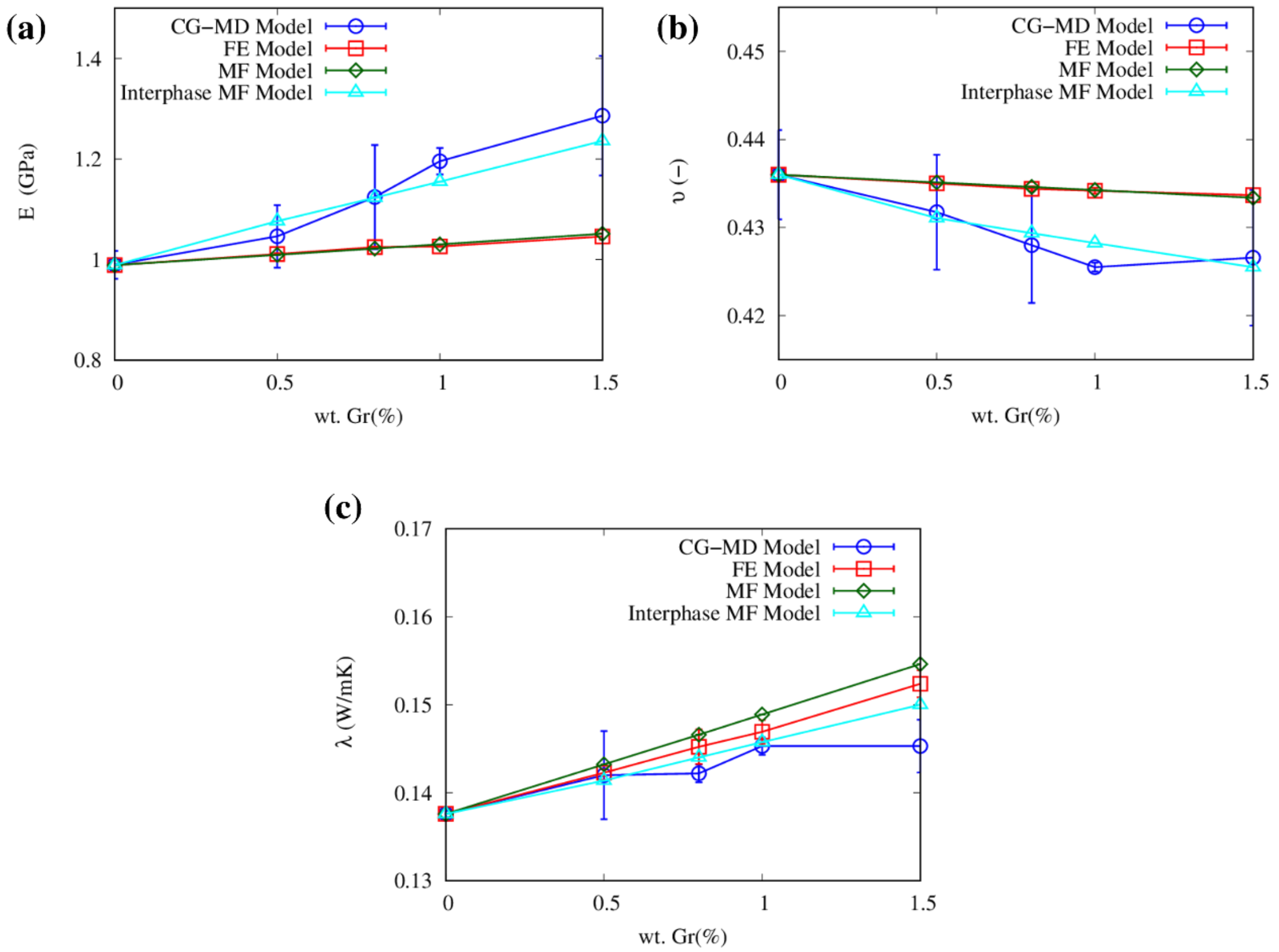

**Figure 10.** Average values with standard deviation of relative (a) Young's modulus, (b) Poisson's ratio, and (c) thermal conductivity enhancement of PP reinforced with graphene. Coarse-Grained (CG), Finite Element (FE), and Mean Field (MF) results (considering either filler and matrix phases, or filler, matrix, and interphase) are compared. Each sample was tested in *x*, *y*, and *z* directions and results averaged.

### *3.5 Multiscale model validation*

We also compared the Young's modulus and thermal conductivity predicted by the calibrated Mori-Tanaka interphase MF model with the experimental results of the tested PP/GNPs nanocomposites. The model input parameters were taken consistently with experimental data, with polypropylene having a density 0.900 g/cm$^3$, Young's modulus 0.944 GPa, Poisson's ratio 0.45, and thermal conductivity 0.23 W/mK. Graphene nanoplatelets (GNPs) were characterized by a density of 2.2 g/cm$^3$, Young's modulus 1,030 GPa, Poisson's ratio 0.19, and thermal conductivity 3,000 W/mK. The interphase properties were taken from the calibration carried out in Section 3.4, leveraging the outcomes of the mesoscopic model.

It is important to note that the aspect ratio of GNPs often deviates significantly from its nominal value upon incorporation into the polymer matrix. For instance, in a study by Kalaitzidou *et al.*,[77] investigating the elastic modulus of xGnP-15/PP composites experimentally, the predicted values from both the Halpin-Tsai and Tandon-Weng models overestimated the experimental results. This discrepancy was attributed to the use of the nominal aspect ratio (~1,500) of xGnP-15 in the calculations, whereas the effective aspect ratio was found to be at least one order of magnitude smaller due to fillers aggregation during composite processing. Similarly, Jun *et al.*[78] examined thermal, mechanical, and electrical properties of PP/GNPs composites containing large-sized GNPs (aspect ratio ~7,500) via melt compounding. Here, the predicted modulus of the composites using the Halpin-Tsai model exceeded the experimental values due to the significantly reduced aspect ratio of GNPs within the composites. However, by employing the aspect ratio measured after composite processing (~50), rather than the nominal aspect ratio, a substantial improvement in agreement between predicted and experimental values was achieved. In another investigation by Innes *et al.*,[79] focusing on rubber matrix reinforced with two types of GNPs (M5 and M15) having lateral dimensions of 5 and 15 μm, with an average thickness in the range of 6-8 nm, micromechanical modeling revealed a relationship between GNP aspect ratio and mechanical properties. The effective aspect ratio contributing to the enhancement of the elastic modulus was determined to be 79 and 86 for M5 and M15 GNPs, respectively, which were significantly smaller than the nominal values.

These findings highlight the importance of adjusting the nominal aspect ratio of GNP fillers after their inclusion into the polymer matrix. Such adjustments are crucial for achieving accurate predictions of the properties in resulting composites. In this study, the Mori-Tanaka interphase MF model showed the best prediction accuracy of elastic modulus and thermal conductivity of experimental PP/GNPs samples (cf. Figure 11) when an effective aspect ratio of 50 was considered, in line with previous studies. [77,78,79]

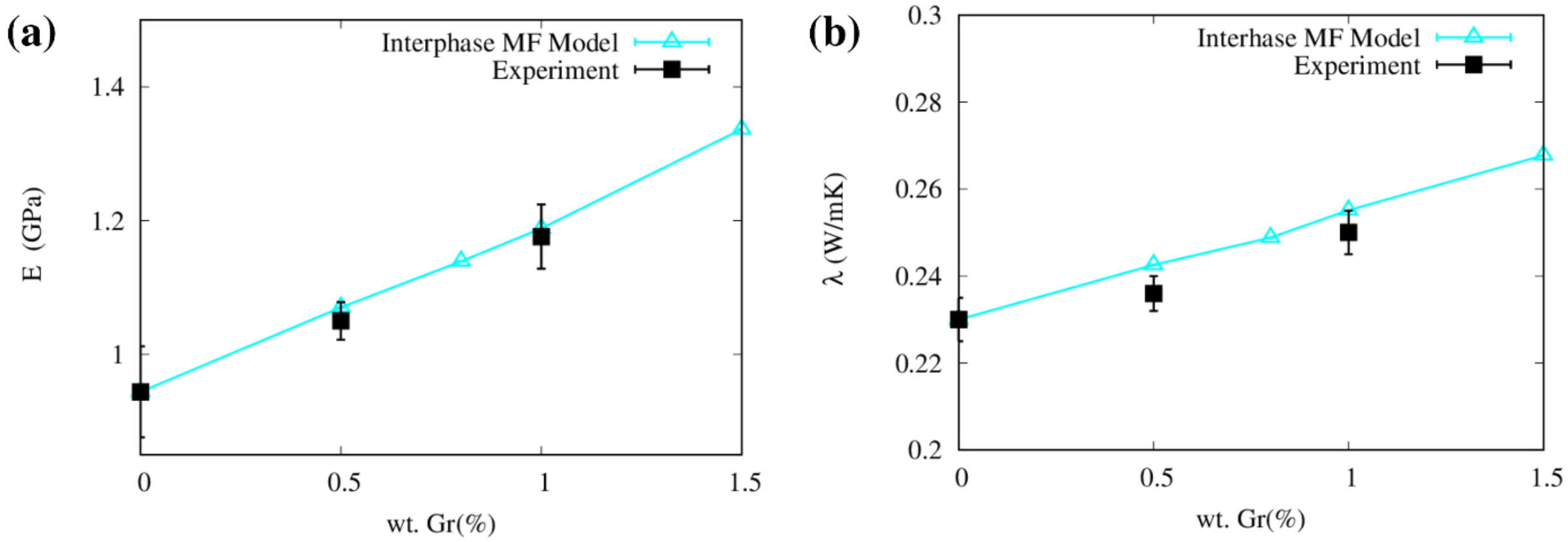

**Figure 11**. Average values with standard deviation of (a) Young's modulus, and (c) thermal conductivity enhancement of PP reinforced with GNPs. Results from experimental measures and interphase MF model predictions are compared (tabulated results are reported in Supplementary Tables S10 and S11).

## 4. CONCLUSION

In this study, we introduce and validate an innovative mesoscopic model for polypropylene nanocomposites reinforced with graphene derivatives, enabling the prediction of their thermal and mechanical properties. The CG-MD simulation method was employed to explore the material properties of both the graphene-based nanofillers and pure polypropylene at the mesoscopic level, with the goal of predicting the properties of the resulting nanocomposite materials.

The mesoscopic simulations revealed that the developed CG model of PP accurately reproduces the glass transition temperature, Young's modulus, Poisson's ratio, and thermal conductivity, exhibiting significant accuracy, except for an overestimation in density. Additionally, we computed the mechanical and thermal properties of graphene using CG potentials, and our results align well with values reported in the literature. Furthermore, we extended our analysis to encompass graphene-oxide and reduced graphene oxide, employing the same CG force-field. This enabled us to predict the degradation in mechanical behavior as the degree of oxidation in graphene-based fillers increases.

The CG model of both polypropylene and graphene-based nanofillers was employed to compute the thermal and mechanical properties of graphene-reinforced polypropylene composites. The CG model predicts that an increase in the nanofiller reinforcement percentage in polypropylene leads to improved mechanical and thermal behavior of the resulting nanocomposite. Our study highlights that the graphene-reinforced polypropylene (PP/Gr) composite exhibits more optimized mechanical properties compared to the composites reinforced with graphene oxide (PP/GO) or reduced graphene oxide (PP/rGO) at similar weight percentages of reinforcement. Conversely, the PP/rGO composite demonstrates superior thermal behavior compared to PP/GO and PP/Gr nanocomposites. To validate the CG predictions and improve prediction accuracy beyond traditional continuum approaches, we utilized the MF model with interphase. This model incorporates calibrated properties based on the results obtained from the mesoscopic model, enabling a detailed representation of the filler-matrix interactions. As a result, the MF model with interphase demonstrated enhanced prediction capability, and its outcomes were validated against experimental data. This validation process corroborated the accuracy of our findings for predicting the thermal and mechanical properties of graphene-reinforced polypropylene nanocomposites.

Our work set the stage for further studies on the multiscale modeling of thermal and mechanical properties of PP nanocomposites reinforced by a broader combination of nanofillers, e.g., carbon nanotubes. Interestingly, our work can have an impact in the energy field. Particularly, additivation of phase-change materials remains an open issue for unlocking the full potential of latent and other heat storage applications.[80,81]

## ASSOCIATED CONTENT

### Supporting Information

The supporting information is available free of charge.

Coarse-grained potentials and tabulated results (PDF)

## ACKNOWLEDGEMENTS

AM, SR, EC, PA and MF thank the CINECA (Iscra C and Iscra B projects) and the Politecnico di Torino's High Performance Computing Initiative (http://hpc.polito.it/) for computing resources. Research scholarship provided to AM by Higher Education Commission (HEC) and Government of Pakistan under title "HRD Initiative-MS leading to Ph.D. program for faculty development for UESTPs/UETs, Phase-I (Batch-V)" for Politecnico di Torino, Italy is also gratefully acknowledged. This work received funding from the European Union's Horizon 2020 research and innovation program, through SMARTFAN (grant agreement no. 760779) project. The authors acknowledge the EsSENce COST Action (CA19118).

## DATA AVAILABILITY

All data generated or analyzed during this study are included in this published article and in the supplementary information file. All simulation files and post-processing codes are available at the Zenodo archive associated to this work (DOI: 10.5281/zenodo.7327415).

# - Supplementary Information -

# Mesoscopic modeling and experimental validation of thermal and mechanical properties of polypropylene nanocomposites reinforced by graphene-based fillers

Atta Muhammad[a,b], Rajat Srivastava[a,c], Nikos Koutroumanis[d], Dionisis Semitekolos[e], Eliodoro Chiavazzo[a], Panagiotis-Nektarios Pappas[d], Costas Galiotis[d,f], Pietro Asinari[a,g], Costas A. Charitidis[e], Matteo Fasano[a*]

[a] *Department of Energy, Politecnico di Torino, Corso Duca degli Abruzzi 24, 10129, Torino, Italy.*

[b] *Department of Mechanical Engineering, Mehran University of Engineering and Technology, SZAB Campus, 66020 Khairpur Mir's, Sindh, Pakistan.*

[c] *Department of Engineering for Innovation, University of Salento, Piazza Tancredi 7, 73100, Lecce, Italy.*

[d] *Institute of Chemical Engineering Sciences, Foundation of Research and Technology-Hellas, Stadiou str. Rion, 26504, Patras, Greece*

[e] *School of Chemical Engineering, National Technical University of Athens, 9 Heroon Polytechniou, 15780 Athens, Greece*

[f] *Department of Chemical Engineering, University of Patras, 1 Caratheodory, 26504 Patras, Greece*

[g] *Istituto Nazionale di Ricerca Metrologica, Strada delle Cacce 91, 10135 Torino, Italy.*

[*] Corresponding author, email address: matteo.fasano@polito.it

SUPPLEMENTARY NOTE 1: COARSE-GRAINED POTENTIALS

i. Polypropylene

| $k_b$ [kcal/mol Å$^2$] | $r_0$ [Å] | $k_\theta$ [kcal/(mol rad$^2$)] | $\theta_0$ [°] | $k_\phi, \phi_s, n$ [kcal/mol][°] [−] |
|---|---|---|---|---|
| 57.36 | 2.98 | 9.32 | 119 | 0.74, 100, 1; -1.41, 190, 2 |

**Table S1**. Parameters of bonded interactions in the PP model.[1]

| σ [Å] | ε [kcal/mol] |
|---|---|
| 4.3 | 0.625 |

**Table S2.** Parameters of non-bonded interactions in the PP model.[1]

ii. Graphene

| Interactions | Parameters |
|---|---|
| Bond | $d_0 = 2.8$ Å<br>$D_0 = 196.38$ kcal/mol<br>$\alpha = 1.55$ Å$^{-1}$<br>$d_{cut} = 3.25$ Å |
| Angle | $\theta_0 = 120^o$<br>$k_\theta = 409.40$ kcal/(mol rad$^2$) |
| Dihedral | $k_\emptyset = 4.15$ kcal/mol; $\phi_s = 0°; n = 2$ |

**Table S3**. Parameters of bonded interactions in the graphene model, where $d_{cut}$ is the bond cutoff (failure criterion) that corresponds to the maximum force on the bond.[2]

| Interactions | Parameters |
|---|---|
| Non-bonded | $\sigma_{lj} = 3.46$ Å<br>$\varepsilon_{lj} = 0.82$ kcal/mol<br>$r_{cut} = 12$ Å |

**Table S4**. Parameters of non-bonded interactions in the graphene model, where $r_{cut}$ is the cutoff radius.[2]

iii. Graphene oxide and reduced graphene oxide

| Interaction | Functional Form |
|---|---|
| Bond | Type I potential and parameters:<br>$V_{b.I}(d) = D_0[\,1 - e^{-\alpha(d-d_0)}]^2$<br>$d_0 = 2.86$ Å<br>$D_0 = 443.07$ kcal/mol<br>$\alpha = 1.54$ Å$^{-1}$<br>$d_{cut} = 3.7$ Å<br>Type II & III:<br>$V_{b,II\&III}(d) = \begin{cases} k_{be}(d-d_0)^2;\ d < d_{c1} \\ k_{bp}(d-d_{c1})^2 + 2k_{be}(d_{c1}-d_0)(d-d_{c1}) + C_1;\ d_{c2} < d < d_{c2} \\ k_{bf}(d-d_{c2})^2 + [2k_{bp}(d_{c2}-d_{c1}) + 2k_{be}(d_{c1}-d_0)](d-d_{c2}) + C_2;\ d > d_{c2} \\ C_1 = k_{be}(d_{c1}-d_0)^2 \\ C_2 = k_{bp}(d_{c2}-d_{c1})^2 + 2k_{be}(d_{c1}-d_0)(d_{c2}-d_1) + C_1 \end{cases}$<br>Type II parameters:<br>$d_0 = 2.94$ Å<br>$d_{c1} = 3.12$ Å<br>$d_{c2} = 3.46$ Å |
| | $k_{be} = 317.34$ kcal/mol Å$^2$<br>$k_{bp} = 126.94$ kcal/mol Å$^2$<br>$k_{bf} = 634.68$ kcal/mol Å$^2$<br>Type III parameters:<br>$d_0 = 2.80$ Å<br>$d_{c1} = 3.00$ Å<br>$d_{c2} = 4.20$ Å<br>$d_{cut} = 4.3$ Å<br>$k_{be} = 256.10$ kcal/mol Å$^2$<br>$k_{bp} = 21.34$ kcal/mol Å$^2$<br>$k_{bf} = 512.20$ kcal/mol Å$^2$ |
| Angle | $V_a(\theta) = k_\theta(\theta - \theta_0)^2$<br>$\theta_0 = 120^0$<br>Type I parameter: $k_\theta = 456.61 \frac{\text{kcal}}{\text{mol}}$<br>Type II parameter: $k_\theta = 259.47 \frac{\text{kcal}}{\text{mol}}$<br>Type III parameter: $k_\theta = 189.93 \frac{\text{kcal}}{\text{mol}}$ |
| Non-bonded | $V_{nb} = 4\varepsilon_{lj}\left[\left(\frac{\sigma_{lj}}{r}\right)^{12} - \left(\frac{\sigma_{lj}}{r}\right)^{6}\right]$<br>$\sigma_{lj} = 7.48$ Å<br>Type C parameter: $\varepsilon_{lj} = 0.0255$ kcal/mol<br>Type H parameter: $\varepsilon_{lj} = 0.128$ kcal/mol<br>Type E parameter: $\varepsilon_{lj} = 0.0797$ kcal/mol |

**Table S5.** Functional forms and parameters of coarse-grained model of graphene oxide and reduced graphene oxide.[3]

SUPPLEMENTARY NOTE 2: TABULATED RESULTS

| wt.% | Young's modulus, E (GPa) | | Thermal conductivity, λ (W/mK) | |
|---|---|---|---|---|
| | PP/GO | PP/rGO | PP/GO | PP/rGO |
| **0.0** | 0.989 (0.027) | 0.989 (0.027) | 0.138 (0.003) | 0.138 (0.003) |
| **0.1** | 1.022 (0.030) | 0.997 (0.134) | 0.143 (0.005) | 0.160 (0.004) |
| **0.5** | 1.054 (0.163) | 1.008 (0.038) | 0.148 (0.001) | 0.163 (0.019) |
| **1.0** | 1.080 (0.059) | 1.052 (0.048) | 0.150 (0.001) | 0.171 (0.009) |
| **1.5** | 1.115 (0.116) | 1.068 (0.197) | 0.151 (0.003) | 0.172 (0.001) |
| **2.0** | 1.119 (0.207) | 1.095 (0.021) | 0.152 (0.001) | 0.173 (0.001) |

**Table S6.** CG-MD results of PP/GO and PP/rGO nanocomposites.

| wt.% | CG MD E (GPa) | Continuum (MF) E (GPa) | Continuum (FE) E (GPa) | Continuum (Interphase MF) E (GPa) |
|---|---|---|---|---|
| **0.0** | 0.989 (0.027) | 0.989 | 0.989(0.027) | 0.989 |
| **0.5** | 1.046 (0.062) | 1.0096 | 1.011(0.001) | 1.076 |
| **0.8** | 1.124 (0.103) | 1.022 | 1.024(0.001) | 1.123 |
| **1.0** | 1.196 (0.026) | 1.0303 | 1.026(0.000) | 1.154 |
| **1.5** | 1.28 (0.118) | 1.0513 | 1.045(0.006) | 1.235 |
| **2.0** | 1.34 (0.134) | 1.0724 | 1.069(0.005) | 1.319 |

**Table S7.** Comparison of Young's modulus obtained from coarse-grained and continuum simulations of PP/Gr.

| wt.% | CG MD Poisson's ratio | Continuum (MF) Poisson's ratio | Continuum (FE) Poisson's ratio | Continuum (Interphase MF) Poisson's ratio |
|---|---|---|---|---|
| **0.0** | 0.436 (0.005) | 0.436 | 0.436 (0.005) | 0.436 |
| **0.5** | 0.431 (0.006) | 0.435 | 0.435 (0.000) | 0.431 |
| **0.8** | 0.428 (0.006) | 0.434 | 0.434 (0.000) | 0.429 |
| **1.0** | 0.425 (0.000) | 0.434 | 0.434 (0.000) | 0.428 |
| **1.5** | 0.426 (0.007) | 0.433 | 0.433 (0.000) | 0.425 |
| **2.0** | 0.427 (0.004) | 0.432 | 0.432 (0.000) | 0.422 |

**Table S8.** Comparison of Poisson's ratio obtained from coarse-grained and continuum simulations of PP/Gr.

| wt.% | CG MD λ (W/mK) | Continuum (MF) λ (W/mK) | Continuum (FE) λ (W/mK) | Continuum (Interphase MF) λ (W/mK) |
|---|---|---|---|---|
| **0.0** | 0.137 (0.000) | 0.137 | 0.137 (0.000) | 0.137 |
| **0.5** | 0.142 (0.008) | 0.143 | 0.142 (0.000) | 0.141 |
| **0.8** | 0.142 (0.000) | 0.146 | 0.145 (0.001) | 0.144 |
| **1.0** | 0.145 (0.002) | 0.148 | 0.146 (0.001) | 0.145 |
| **1.5** | 0.145 (0.000) | 0.155 | 0.152 (0.001) | 0.15 |
| **2.0** | 0.144 (0.003) | 0.160 | 0.155 (0.002) | 0.154 |

**Table S9.** Comparison of thermal conductivity obtained from coarse-grained and continuum simulations PP/Gr.

| wt.% | Continuum (Interphase MF) E (GPa) | Experimental E (GPA) |
|---|---|---|
| **0.0** | 0.944 | 0.944 (0.068) |
| **0.5** | 0.107 | 1.05 (0.028) |
| **0.8** | 1.139 | - |
| **1.0** | 1.188 | 1.176 (0.048) |
| **1.5** | 1.336 | - |

**Table S10.** Comparison of Young's modulus obtained from experiments and continuum simulations of PP/GNPs.

| wt.% | Continuum (Interphase MF) λ (W/mK) | Experimental λ (W/mK) |
|---|---|---|
| **0.0** | 0.23 | 0.23 (0.005) |
| **0.5** | 0.242 | 0.236 (0.003) |
| **0.8** | 0.248 | - |
| **1.0** | 0.255 | 0.250 (0.005) |
| **1.5** | 0.267 | - |

**Table S11.** Comparison of thermal conductivity obtained from experiments and continuum simulations PP/GNPs.